\documentclass[10pts,showpacs,preprint,aps]{revtex4}
\usepackage{graphicx}
\usepackage{dcolumn}
\usepackage{bm}
\usepackage{amssymb}
\usepackage{amsmath}
\begin{document}
\setcounter{page}{1}
\vskip 2cm
\title 
{Relative locality and relative Co-locality as extensions of the Generalized Uncertainty Principle}
\author
{Ivan Arraut}
\affiliation{  
Department of Physics, Osaka University, Toyonaka, Osaka 560-0043, Japan}

\begin{abstract}
I introduce the spatial curvature effects inside the formalism of Relative Locality as a non-commutative structure of the momentum space in agreement with the very well known concepts of Quantum Groups. This gives a natural red-shift effect in agreement with an extended version of the Generalized Uncertainty Principle (GUP) and in agreement with the standard notions of curvature inside General Relativity. I then introduce the concept of Relative Co-locality as a reinterpretation of the usual notions of spacetime curvature. From this redefinition, I obtain the momentum space curvature effects as a non-commutativity in position space. This introduce a natural blue-shift effect in agreement with the extended version of GUP.  
Both effects, Relative locality and Co-locality are dual each other inside the formalism of quantum groups $SU_q(n)$ symmetric Heisenberg algebras and their q-Bargmann Fock representations. When Relative locality and Co-locality are introduced, the q-deformation parameter takes the form $q\approx 1+\sqrt{\frac{\vert p\vert \vert x\vert}{r_\Lambda m_{pl}}}$ with the spatial curvature effects in Relative Locality appearing like $\Delta X\approx \frac{\vert x\vert}{m_{pl}}\Delta P$ and the momentum curvature effects in Relative Co-locality appearing like $\Delta P\approx \frac{\vert p\vert}{r_\Lambda}\Delta X$, where $r_\Lambda=\frac{1}{\sqrt{\Lambda}}$ is the scale defined by the Cosmological Constant $\Lambda$, $m_{pl}$ is the Planck mass and $\Delta X/\Delta P$ is a scale of position/momentum or time/energy associated with the event, p and x are the momentum and position of the observer relative to the event. 
\end{abstract}
\maketitle 
\section{Introduction}
Every approach of Quantum Gravity agrees in the fact that the spacetime becomes discrete at the Planck scale. It has been recently argued that it is not necessary to go to the Planck scale in order to perceive the effects of Quantum Gravity \cite{GAC, 1, 2}. In fact, the Gamma ray bursts and the possibility of observing some energy dependent delays of arrival times of photons produced simultaneously at a given distance offers a good possibility for testing the geometry of momentum space. This is the so-called principle of Relative Locality introduced by Giovanni Amelino-Camelia, Laurent Freidel, Jerzy Kowalski-Glikman and Lee Smolin \cite{1,2}. In such a case, the Born principle already formulated in 1.938 is taken seriously \cite{2}. The Born principle states that in the quantum gravity regime, momentum and position must be taken on equal footing, in other words there must be a symmetry between momentum and position in the same sense that there is a symmetry between space and time in special relativity for example. Relative locality is an extension of the Relativity principle to the phase space, this principle suggests that events that are local with respect to one observer, are not necessarily local with respect to an observer at a different position. The observers basically live in different spacetimes created as the cotangent bundle of the momentum space \cite{S1}. The fundamental scale of the theory is the planck mass $m_{pl}$. Then some physical effects can be observed at astrophysical scales for example. The regime of Relative locality is $G_N\to0$ and $\hbar\to0$ but keeping the scale $m_{pl}=\sqrt{\frac{\hbar}{G_N}}$ constant. 
Still is open the question of how can we include the spacetime curvature effects inside this formalism and how can be introduced the Cosmological Constant $\Lambda$ scale. There is a proposal due to Giovanni Amelino-Camelia, Antonio Marciano, Marco Tanassa and Giacomo Rosati \cite{GAC2}.
Here I take an alternative path by extending the notions of quantum groups $SU_q(n)$ symmetric Heisenberg algebras and their q-Bargmann Fock representations. In Relative Locality, if the momentum space is assumed to be non-commutative at some scale given by the space curvature, then it is possible to derive the gravitational red-shift effect if we interpret appropriately GUP.
I then postulate the concept of "Relative Co-locality" which is just a reinterpretation of the usual notions of spacetime curvature. Here however, we can introduce the momentum curvature as a non-commutative effect of the space in some scale. The curvature effects for Relative Locality appear like $\Delta X\approx \frac{\vert x\vert}{m_{pl}}\Delta P$, where $\Delta P$ is the scale of momentum or energy related to the event. It is a consequence of the non-commutativity of momentum space. On the other hand, the momentum curvature effects in Relative Co-locality appear like $\Delta P\approx \frac{\vert p\vert}{r_\Lambda}\Delta X$ with $r_\Lambda$ fixing the radius of curvature of the position space.
The Relative Co-locality expression is just the dual version of the Relative Locality one in agreement with the $SU_q(n)$ symmetry with a q-deformation parameter $q\approx 1+\sqrt{\frac{\vert p\vert \vert x\vert}{r_\Lambda m_{pl}}}$. This is just an extension of the deformation parameter q when we take $l_{pl}$ as the minimum scale in position and $\frac{1}{r_\Lambda}$ as the minimum scale in momentum \cite{4}.
In a general process or event, the Relative locality effects must compete with the Relative co-locality ones. They reproduce opposite contributions in agreement with a physical observer moving with respect to any event.
The paper is organized as follows: In Section \ref{eq:S1}, I make a review of the standard (Lagrangian) formalism of Relative Locality. In Section \ref{eq:dS BH}, I introduce the curvature effects inside the Relative Locality formalism by giving a non-commutative structure for the momentum space in agreement with the standard notions of quantum groups, there is a natural variation of the UV cut-off scale with respect to the observer's position. In Section \ref{eq:8}, I make a review of GUP with UV and IR cut-offs inside the q-Bargmann Fock formalism already suggested by Kempf with some extensions. In Section \ref{eq:RLoc}, I introduce the Relative Co-locality formalism in order to introduce the momentum curvature effects as a non-commutativity of position. In Section \ref{eq:RCL}, I introduce both, Relative Locality and Co-locality as extensions of GUP. Here the scale $\Lambda$ appears and it restores the $SU_q(n)$ symmetry, necessary inside the q-Bargmann Fock formalism. In Section \ref{eq:New}, I show how can we isolate the curvature effects in position by fixing the observer's momentum scale. Finally in Section \ref{eq:C}, I conclude.  

\section{The action principle for Relative Locality}    \label{eq:S1}

First I will make a brief review of the concepts of Relative Locality already introduced in the literature \cite{1,2} in agreement with its action principle. The action for a particle moving is in general given by:

\begin{equation}   \label{eq:1a}
S=\sum_{world lines,I}S^I_{free}+\sum_{interaction,\alpha}S^\alpha_{int}
\end{equation} 

where $S_{free}$ is the action for a free particle and $S_{int}$ is the action representing the interaction vertex of the point particle \cite{1,2}. The action for the free particle is:

\begin{equation}   \label{eq:2a}
S^I_{free}=\int ds(x_I^a\dot{p}^I_a+N_I C^I(p^I))
\end{equation}

where $s$ is an arbitrary time parameter and $N_I$ is a Lagrange multiplier imposing the on-shell condition:

\begin{equation}   \label{eq:3a}
C^I(k)\equiv D^2(p)-m^2_I
\end{equation}

$D^2(k)$ is the geodesic distance from the origin of momentum space to the point denoted by the momentum $k_a$. The interaction is described by:

\begin{equation}   \label{eq:4a}
S^\alpha_{int}=-K_a^{(\alpha)}z^a_{(\alpha)}
\end{equation}

the equations of motion can then be derived as:

\begin{equation}   \label{eq:5a}
\dot{p}_a^I=0
\end{equation}

\begin{equation}   \label{eq:6a}
\dot{x}^a_I=N_I\frac{\delta C^I}{\delta p_a^I}
\end{equation}

\begin{equation}   \label{eq:7a}
C^I(p)=0
\end{equation}

\begin{equation}   \label{eq:8a}
K_a^{(s)}=0
\end{equation}

this last expression corresponds to the momentum conservation at each vertex. On the other hand, the most interesting relation is obtained by doing the variation with respect to $p_a(0)$ at the end points of the world line in equations \ref{eq:2a} and \ref{eq:4a}. The obtained expression is then given by:

\begin{equation}   \label{eq:9a}
x^a_I=\pm z^b\frac{\delta K_b}{\delta p_a}
\end{equation}

this is the Relative locality expression derived from the action principle. The positive sign applies for incoming momenta and the negative sign appears for outgoing momenta. 

\section{Curvature effects in Relative Locality as a non-commutative effect of the momentum space}   \label{eq:dS BH}

In Relative Locality the events are (almost) local with respect to the observers near them, but non-local with respect to observers far away from it. This is a consequence of the fact that different observers construct different spacetimes as a cotangent bundle over a curved momentum space \cite{1,2}. 
Then if we have a collision at a given point in space for example, in principle an observer very far from the event will say that this collision was due to a non-local interaction or he/she will describe it as a non-local event. By interaction I mean the collision itself for example. How do you describe the collision of particles very far from the event? How do you describe QFT and how can you establish the QFT rules inside this formalism is a good challenge which can be solved if we use the appropriate formulation. From eq. \ref{eq:9a}, the non-local correction due to Relative Locality is given by \cite{1,2} ($c=1$):

\begin{equation}   \label{eq:1}
\delta X\approx \vert x\vert\frac{E}{m_{pl}}
\end{equation}
   
where E is the energy associated with the event; $m_{pl}$ is the Planck mass and x is the relative position between the observer and the event. The Planck length is not the fundamental scale in Relative locality. The Relative Locality regime is $\hbar\to0$, which means a classical regime and $G_N\to0$, which means the absence of gravity. But keeping $m_{pl}=\sqrt{\frac{\hbar}{G_N}}$ fixed. Here I will introduce the curvature effects as a non-commutativity of momentum space. For that purpose we have to keep in mind that the curvature in position space can be introduced as a non-commutativity of momentum space. Let's start with the Classical Poisson brackets defined as:

\begin{equation}   \label{eq:3pp}
\{x^a,p_b\}=\delta^a_{\;\;b}
\end{equation}

all other Poisson brackets are zero. Suppose now that we have many particles interacting on a common vertex, the momentum associated with the system is then given by $P_{tot}=\sum_i p_i$. Then the translation for the world lines for each particle is given by:

\begin{equation}   \label{eq:4pp}
\delta x^\mu_I=\{x^\mu_I,b^\nu P_\nu^{Tot}\}=\{x^\mu_I,b^\nu\sum_J p_\nu^J\}=b^\mu
\end{equation}   

then all the world lines are translated together, this is independent of the momentum of each particle. This is just the notion of absolute locality taken from Special Relativity \cite{2,3}. If we consider now the Relative Locality correction with the momentum space having a non-trivial connection, then in agreement with \cite{2,3}, the momentum superposition is non-linear and given by:

\begin{equation}   \label{eq:5pp}
P_\mu^{tot}=\sum_I p^I_\mu+\frac{1}{m_{pl}}\sum_{I<J}\Gamma^{\nu\rho}_\mu p_\nu^Ip_\rho^J
\end{equation}

and then the translation for the world lines in agreement with the Poisson brackets, become:

\begin{equation}   \label{eq:6pp}
\delta x_I^\mu=\{x^\mu_I, b^\nu P_\nu^{tot}\}=b^\mu+\frac{1}{m_{pl}}b^\nu\sum_{J>I}\Gamma^{\mu\rho}_\nu p_\rho^J
\end{equation}

in this case, the translation of each world line depends on the momentum carried by the particle and on the other particles which interact with it. At the quantum level, the Poisson brackets become to be commutators, in such a case, if we assume only minimal scale in position (UV cut-off), then the commutation relation between $x$ and $p$ becomes \cite{55}:

\begin{equation}   \label{eq:7ppepe}
[x,p]=i\hbar(1+\beta p^2+...) 
\end{equation}

then GUP with the UV cut-off becomes \cite{3,4}:

\begin{equation}   \label{eq:2}
\Delta X\Delta P\geq \frac{1}{2}\left(\hbar+\frac{l_{pl}^2}{\hbar}(\Delta P)^2\right)
\end{equation}
 
it is possible to extend this result in order to include the notions of Relative Locality if we take into account the non-commutative structure of position space. In ref. \cite{1}, the Relative Locality expression is equivalent to:

\begin{equation}   \label{eq:ARR1}
x^\mu_{\;\;B}=[U_\gamma]^\mu_{\;\;\nu}x^\nu_A
\end{equation}  

with the linear transformation defined as:

\begin{equation}   \label{eq:ARR2}
[U_\gamma]^\mu_{\;\;\nu}=\delta^\mu_{\;\;\nu}+\frac{1}{m_{pl}}\Gamma^{\mu\rho}_\nu p_\rho
\end{equation}

where $\Gamma^{\mu\rho}_\nu$ is the connection defined in a curved momentum space. By introducing this result in eq. \ref{eq:ARR1} and then performing the commutation with respect to the variable $p$, the we get:

\begin{equation}   \label{eq:ARR3}
[x^\mu_B,p_\gamma]=[x^\mu_A,p_\gamma]+\frac{\Gamma^{\mu\rho}_\nu}{m_{pl}}\left(p_\rho[x^\nu_A,p_\gamma]+[p_\rho,p_\gamma]x^\nu_A\right)
\end{equation}

as the standard version of Relative locality is performed over a classical configuration of space, then it is perfectly valid to say that at first quantization the relation $[x,p]=i\hbar$ is valid. Here we will assume a non-commutativity structure for the momentum space in the form introduced in \cite{Majid}, for that purpose we will restrict the analysis for 3 Dimensions. The non-commutative in momentum space will be introduced like $[p_i,p_j]=\frac{i}{R}\epsilon_{ijk}p_k$, where $R$ is just the radius of curvature in space. It could be perfectly taken as $r_\Lambda$, the scale of the de-Sitter space. But in principle it could be arbitrary. Expression \ref{eq:ARR3} then becomes:

\begin{equation}   \label{eq:ARR4}
[x,p]=i\hbar\left(1+\frac{\Gamma^{\mu\rho}_\mu p_\rho}{m_{pl}}+\frac{\Gamma^{\mu\rho}_\nu[p_\rho,p_\mu]x^\nu_A}{i\hbar m_{pl}}\right)
\end{equation} 

if we apply the standard formula of Quantum Mechanics given by:

\begin{equation}   \label{eq:ARR5}
\Delta X\Delta P\geq\frac{1}{2}\vert<[x,p]>\vert
\end{equation}

then, the expression \ref{eq:ARR4} is perfectly equivalent to:

\begin{equation}   \label{eq:ARR6}
\Delta X\Delta P\geq \frac{\hbar}{2}\left(1+\left<\frac{\Gamma^{\mu\rho}_\mu p_\rho}{m_{pl}}\right>+\left<\frac{\Gamma^{\mu\rho}_\nu \epsilon_{\rho\mu\kappa}p_\kappa}{\hbar r_\Lambda m_{pl}}\right>x^\nu\right)
\end{equation}

where $\epsilon_{\mu\rho\kappa}$ is the 3D Levi-Civita symbol just introduced in order to mimic the non commutative behavior of the momentum space and $r_\Lambda$ has been introduced as the radius of curvature in space. Note the in eq. \ref{eq:ARR6}, we allow the connection to have a torsion contribution. Note also that we assume the expectation value not to affect the distance variable for the third term on the right-hand side, in fact, in that term, $x^\nu$ can be assumed to be classical if the distance is much smaller than the scale $r_\Lambda$. Then in agreement with eq. \ref{eq:ARR5}, we can trust in the following equality:

\begin{equation}
(\Delta P)^2\approx\frac{1}{2}\vert<[p_i,p_j]>\vert=\left<\frac{\Gamma^{\mu\rho}_\nu \epsilon_{\rho\mu\kappa}p_\kappa}{r_\Lambda}\right>
\end{equation}

if we assume now that we are analyzing some event with a total momentum near zero, i.e Center of mass frame, In such a case eq. \ref{eq:ARR6} becomes:

\begin{equation}   \label{eq:ARR7}
\Delta X\Delta P\geq \frac{\hbar}{2}\left(1+\frac{\vert x\vert}{\hbar m_{pl}}(\Delta P)^2\right)
\end{equation}

then we have an additional uncertainty given by:

\begin{equation}   \label{eq:4}
\Delta X \Delta P \approx \vert x\vert \frac{(\Delta P)^2}{m_{pl}}
\end{equation}

if we compare with the UV cut-off for GUP given by eq. \ref{eq:1}, we observe that both expressions are comparable if ($\hbar=1)$:

\begin{equation}   \label{eq:5}
l_{pl}^2=\frac{\vert x\vert}{m_{pl}}\;\;\;\;\;\to \vert x\vert=l_{pl}^2m_{pl}
\end{equation}

the UV cut-off is just the standard one if the observer is at a distance equivalent to the Planck length $x=l_{pl}$. Only an observer at this distance will perceive the events to be local and he will describe GUP in agreement with the standard version. We can rewrite the previous expression as:

\begin{equation}   \label{eq:6}
\Delta X\Delta P\geq \frac{1}{2}\left(1+\frac{\vert x\vert}{m_{pl}}(\Delta P)^2\right)
\end{equation} 

in such a case, the UV cut-off is modified to $\Delta X_{min}=\sqrt{\frac{\vert x\vert}{m_{pl}}}$ and $\Delta P_{UV}=\sqrt{\frac{m_{pl} }{\vert x\vert}}$, the subindex UV means "Ultraviolet". The interpretation is then clear. An observer very far from the source, will perceive a larger minimum scale in position (increase of non-locality) and as a consequence, an smaller UV scale in momentum. The UV cut-off moves to the IR as the observer is far from the source. This is a natural red-shift effect and it is a consequence of Relative Locality in momentum space, where the non-commutativity of the momentum space simulates the spatial curvature effects. 
In other words, an observer near enough to the UV collision, will say that the cut-off is $l_{pl}$, on the other hand, an observer at a distance $\vert x\vert$, will say that the cut-off is moved to the IR in agreement with the expression \ref{eq:6}. However, both observers would agree with a formulation of GUP in agreement with their experiences. They both live in different spaces, but in the same phase space. Let's assume now that the largest distance between the observer and the source is given by the Cosmological Constant $\Lambda$ scale. For an observer at $r=r_\Lambda$, the GUP expression would be:

\begin{equation}   \label{eq:7}
\Delta X\Delta P\geq \frac{1}{2}\left(1+\frac{r_\Lambda}{m_{pl}}(\Delta P)^2\right)
\end{equation}
 
with the corresponding UV cut-off given by $\Delta X_{min}=\sqrt{l_{pl}r_\Lambda}$ and $\Delta P_{max}=\sqrt{\frac{1}{l_{pl}r_\Lambda}}$. This is an interesting result because this is the UV-IR mix scale already derived as a general extremal condition in \cite{4} when $l_{pl}$ is an UV cut-off and $r_\Lambda$ is an IR one. In the present case, we see that this scale is the maximum degree of red-shift due to the curvature effects inside the Relative Locality formalism.

\section{GUP with UV and IR cut-off scales: A Review}   \label{eq:8}

It is useful at this point to make a review of GUP inside the q-Bargmann Fock formalism. In agreement with Kempf \cite{5}, we can define the Bargmann Fock operators:

\begin{equation}   \label{eq:9}
\bar{\eta}:=\frac{1}{2L}x-\frac{i}{2K}p\;\;\;\;\;\partial_{\bar{\eta}}:=\frac{1}{2L}x+\frac{i}{2K}p
\end{equation} 

where the constants L and K carry units of length and momentum. The commutation relations are then generalized to $\partial_{\bar{\eta}}\bar{\eta}-q^2\bar{\eta}\partial_{\bar{\eta}}=1$; with $q\geq1$. q is related to the gravitational degrees of freedom. This formalism has 2 free parameters with the constants K and L are related by $KL=\frac{1}{4}\hbar(q^2+1)$. The commutation relations of the position and momentum operator are now given by:

\begin{equation}
[x,p]=i\hbar+i\hbar(q^2-1)\left(\frac{x^2}{4L^2}+\frac{p^2}{4K^2}\right)
\end{equation}

from this follows the uncertainty relation\cite{5}:

\begin{equation}   \label{eq:10}
\Delta X \Delta P \geq \frac{\hbar}{2}\left(1+(q^2-1)\left(\frac{(\Delta X)^2}{4L^2}+\frac{(\Delta P)^2}{4K^2}\right)\right)
\end{equation}

here we have assumed $<X>=<P>=0$ \cite{4,5}. In agreement with the formalism developed in \cite{4,5}, the minimum scale in position corresponds to:

\begin{equation}   \label{eq:26}
\Delta X_{min}=L\sqrt{1-q^{-2}}
\end{equation}  

additionally, it is known that the smallest uncertainty in momentum is given by:

\begin{equation}   \label{eq:27}
\Delta P_{min}=K\sqrt{1-q^{-2}}
\end{equation}

K and L must satisfy the additional constraint:

\begin{equation}   \label{eq:28}
KL=\frac{(q^2+1)\hbar}{4}
\end{equation}

this condition suggests that the UV scale is dual to the IR one. For the moment, as a fact of illustration, the two free parameters will be fixed in agreement with the Planck scale $l_{pl}$ and the Cosmological Constant scale $\Lambda=\frac{1}{r_\Lambda^2}$. We want an expression for GUP symmetric with respect to position and momentum given by:

\begin{equation}   \label{eq:29}
\Delta X \Delta P\geq \frac{\hbar}{2}+\frac{l_{pl}^2}{2\hbar}(\Delta P)^2+\frac{\hbar}{2r_\Lambda^2}(\Delta X)^2
\end{equation}

if we want this expression to agree with \ref{eq:10}, the following conditions must be satisfied:

\begin{equation}   \label{eq:30}
l_{pl}=L\sqrt{1-q^{-2}}   
\end{equation}

\begin{equation*}   
\frac{1}{r_\Lambda}=K\sqrt{1-q^{-2}}
\end{equation*}

these expressions together with the condition \ref{eq:28}, automatically fix the value of q. Although q could take different values in order to satisfy the previous conditions, we select the value of q satisfying the additional constraint $q\geq 1$ \cite{5}. In such a case, q, which is related to the gravitational degrees of freedom, satisfies:

\begin{equation}   \label{eq:31}
q\approx 1+\frac{l_{pl}}{r_\Lambda}
\end{equation}

where $l_{pl}\approx 10^{-35}mt$ is the Planck scale and $r_\Lambda\approx 10^{26} mt$ is the Hubble one, then $q\approx 1+10^{-61}$. If $\Lambda\to0$ then $q=1$ and $l_{pl}\to0$ and vice versa. It means that inside this formalism, the Cosmological Constant $\Lambda$ is related to the minimum scale in position. Without a minimum scale, there is no Cosmological Constant and vice versa.
Originally Kempf \cite{4,5} derived the results \ref{eq:26} and \ref{eq:27} by defining the function:

\begin{equation}   \label{eq:111}
f(\Delta X, \Delta P):=\Delta X \Delta P-\frac{\hbar}{2}\left(1+(q^2-1)\left(\frac{(\Delta X)^2+<X>^2}{4L^2}+\frac{(\Delta P)^2+<P>^2}{4K^2}\right)\right)
\end{equation}  

here, we will assume $<X>=0=<P>$. The minimum scale in position can be found given the following extremal condition:

\begin{equation}   \label{eq:112}
\frac{\partial}{\partial \Delta P}f(\Delta X, \Delta P)=0\;\;\;\;\;f(\Delta X, \Delta P)=0
\end{equation}

then, the result is just the eq. \ref{eq:30} $\left(\Delta X_{min}=L\sqrt{1-q^{-2}}\right)$. On the other hand, the minimum scale in momentum is obtained from the condition:

\begin{equation}   \label{eq:113}
\frac{\partial}{\partial \Delta X}f(\Delta X, \Delta P)=0\;\;\;\;\;f(\Delta X, \Delta P)=0
\end{equation}

the result is just the eq. \ref{eq:30} $\left(\Delta P_{min}=K\sqrt{1-q^{-2}}\right)$. We can however, derive a third scale given by the UV-IR mix effects. This scale was introduced for first time by John A. Wheeler \cite{W} in 1957. It is given by the geometric average of the $l_{pl}$ and $r_\Lambda$, namely, $l_0=(l_{pl}r_\Lambda)^{1/2}$.
We can define the total differential for the function $f(\Delta X, \Delta P)$ as:

\begin{equation}   \label{eq:114}
df(\Delta X, \Delta P)=\left(\frac{\partial f(\Delta X, \Delta P)}{\partial \Delta P}\right)_{\Delta X=C}d(\Delta P)+\left(\frac{\partial f(\Delta X, \Delta P)}{\partial \Delta X}\right)_{\Delta P=C}d(\Delta X)
\end{equation}

The general extremal condition inside the phase space is obtained as the total differential \ref{eq:114} goes to zero. In such a case:

\begin{equation}   \label{eq:115}
df(\Delta X, \Delta P)=0
\end{equation}

obtaining then the result:

\begin{equation}   \label{eq:116}
\frac{d(\Delta X)}{d(\Delta P)}=-\frac{\Delta X}{\Delta P}\frac{\left(1-\frac{\hbar}{4K^2}(q^2-1)\frac{\Delta P}{\Delta X}\right)}{\left(1-\frac{\hbar}{4L^2}(q^2-1)\frac{\Delta X}{\Delta P}\right)}
\end{equation}

imposing then the additional condition $\frac{d(\Delta X \Delta P)}{d\Delta X}=0=\frac{d(\Delta X \Delta P)}{d\Delta P}$. Then, we have to satisfy:

\begin{equation}   \label{eq:121}
\frac{\Delta X}{\Delta P}=-\frac{d(\Delta X)}{d(\Delta P)}
\end{equation}

introducing this result inside the right-hand side of \ref{eq:116}, we obtain:

\begin{equation}   \label{eq:122}
(q^2-1)\left(\frac{\Delta X}{L^2}+\frac{\Delta P}{K^2}\frac{d(\Delta P)}{d(\Delta X)}\right)=0
\end{equation}

this equation has two solutions. The first one is not interesting for us, because it suggests $q=1$ which is a trivial condition because in such a case $d(\Delta X \Delta P)=0$ everywhere. Additionally, $q=1$ corresponds to the standard Bosonic algebra in agreement with \cite{1, 4, 5}. We then do not consider that case here. The interesting case is:

\begin{equation}   \label{eq:123}
\left(\frac{\Delta X}{L^2}+\frac{\Delta P}{K^2}\frac{d(\Delta P)}{d(\Delta X)}\right)=0
\end{equation}   

which in combination with \ref{eq:121} gives:

\begin{equation}   \label{eq:124}
\Delta X=\pm \frac{L}{K}\Delta P
\end{equation}

if we compare the expressions \ref{eq:29} with the one obtained in \ref{eq:10}, then it is simple to verify that ($\hbar=1$):

\begin{equation}   \label{eq:125}
L=K^{-1}=\frac{\sqrt{2}}{2}(l_{pl}r_\Lambda)^{1/2}
\end{equation}

then the condition \ref{eq:124} becomes:

\begin{equation}   \label{eq:126}
\Delta X=\frac{1}{2}(l_{pl}r_\Lambda)\Delta P
\end{equation}

here we have only taken into account the positive sign. For consistence, we can verify that the condition \ref{eq:28} is satisfied. If we take into account that in agreement with \ref{eq:31}, we have $q^2\approx 1+2\frac{l_{pl}}{r_\Lambda}$. Then \ref{eq:28} becomes ($\hbar=1$):

\begin{equation}   \label{eq:127}
KL\approx\frac{1}{2}
\end{equation} 

This result is consistent with \ref{eq:125} as can be verified. If we replace \ref{eq:126} inside \ref{eq:29}, we then obtain under the approximation $r_\Lambda>>l_{pl}$, the following result:

\begin{equation}   \label{eq:128}
\Delta X\approx (l_{pl}r_\Lambda)^{1/2}=l_0\;\;\;\;\;\Delta P\approx \frac{1}{(l_{pl}r_\Lambda)^{1/2}}=\frac{1}{l_0}
\end{equation}

The result \ref{eq:128} is just the UV-IR scale already suggested by John A. Wheeler in 1957 \cite{W} and interpreted as a coherence region. Note that this is also the maximum possible degree of non-locality in momentum space in agreement with the expressions obtained after eq. \ref{eq:7}. In such a case, the $\Lambda$ scale was introduced artificially as the maximum possible distance for an observer relative to some event.

\section{Relative Co-Locality: The standard notion of curvature in space}   \label{eq:RLoc}   

Relative Co-locality is obtained by just introducing the spatial curvature in the standard way. In agreement with Shan Majid \cite{Majid}, a non commutativity in space is equivalent to a curvature in momentum space and a non-commutativity in momentum is just equivalent to a curvature in position. Under this philosophy, the concept of Relative Locality introduced by Amelino-Camelia and colleagues is incomplete since the non-commutative nature of the momentum coordinates is not included in its standard version given by $[p_i,p_j]\neq0$. On the other hand, in the standard notion of a curved spacetime, if we want to compare the momentum of two particles located at different positions, we have to parallel transport by using an affine connection. Then the result is:

\begin{equation}   \label{eq:lala}
p_\nu^B=[U_\gamma]^\mu_{\;\;\nu} p_\mu^A
\end{equation}  

with:

\begin{equation}   \label{eq:lala2}
[U_\gamma]^\mu_{\;\;\nu}=\delta ^\mu_{\;\;\nu}+\frac{1}{r_\Lambda}\Gamma^\mu_{\nu\beta}x^\beta
\end{equation}

if we commute the expression \ref{eq:lala} with $x^\gamma$, we get:

\begin{equation}   \label{eq:lala3}
[x^\gamma,p_\nu^B]=[x^\gamma,p_\nu^A]+\frac{\Gamma^\mu_{\nu\beta}}{r_\Lambda}\left(x^\beta[x^\gamma,p_\mu^A]+[x^\gamma,x^\beta]p_\mu^A\right)
\end{equation}
  
if we introduce the standard notions of uncertainty and if additionally, we take into account the non-commutativity of space given by:

\begin{equation}
[x_i,x_j]=\frac{i}{m}\epsilon_{ijk}x_k
\end{equation}

where $m$ is the curvature associated with the momentum space, then eq. \ref{eq:lala3} becomes:

\begin{equation}   \label{eq:lala4}
[x,p]=i\hbar\left(1+\frac{\Gamma^\nu_{\nu\beta}}{r_\Lambda}x^\beta+\frac{\Gamma^\mu_{\nu\beta}}{\hbar m_{pl}r_\Lambda}\epsilon^{\nu\beta\kappa}x^\kappa p_\mu\right)
\end{equation}

where in this case, we take the curvature of momentum space as equivalent to the Planck mass as an observer in position space would describe. By applying the same formula as in eq. \ref{eq:ARR5}, then we get:

\begin{equation}   \label{eq:lala5}
\Delta X\Delta P\geq \frac{\hbar}{2}\left(1+\left<\frac{\Gamma^\nu_{\nu\beta}x^\beta}{r_\Lambda}\right>+\left<\frac{\Gamma^\mu_{\nu\beta}}{\hbar m_{pl}r_\Lambda}\epsilon^{\nu\beta\kappa}x^\kappa\right> p_\mu\right)
\end{equation}

and taking into account that:

\begin{equation}  
(\Delta X)^2\approx \left<\frac{\Gamma^\mu_{\nu\beta}}{m_{pl}}\epsilon^{\nu\beta\kappa}x^\kappa\right>
\end{equation}

and if we analyze the system around the center of mass, eq. \ref{eq:lala5} becomes:

\begin{equation}   \label{eq:lala6}
\Delta X\Delta P\geq \frac{\hbar}{2}\left(1+\frac{\vert p\vert}{\hbar r_\Lambda}(\Delta X)^2\right)
\end{equation}

note that again we allow the connection to have some torsion contribution in eq. \ref{eq:lala5}. The combined results of this section and Section \ref{eq:dS BH} are introduced in the next one with the aid of the q-Bargmann Fock formalism. In such a case, the parameter q, takes some specific form in order to satisfy the previous results. 

\section{Relative Locality and Co-locality as extensions of GUP}   \label{eq:RCL}

Here we introduce the principle of Relative Locality and Co-locality together as extensions of GUP. This is one way to introduce the notions of Relative locality in a curved spacetime or Relative Co-locality in curved momentum space, the curvature space is just a manifestation of the non-commutativity of momentum and the curvature in momentum space becomes a manifestation of the non-commutativity in position. In this way, we make a full extension of the Born principle as has been explained previously. The curvature scales are in principle arbitrary but they are related through some constraint as has already been explained. The expression which provides consistence with the $SU_q(n)$ symmetric formulation inside a q-Bargmann Fock formalism is:

\begin{equation}   \label{eq:11}
\Delta X \Delta P\geq \frac{1}{2}\left(1+\frac{\vert x\vert}{ m_{pl}}(\Delta P)^2+\frac{\vert p\vert}{ r_\Lambda}(\Delta X)^2\right)
\end{equation}

where $\hbar=1$. Note that the IR cut-off becomes the standard one given in eq. \ref{eq:29} as $\vert p\vert=\frac{1}{r_\Lambda}$, which is just the minimum momentum for the observer relative to the events. If we define the function:

\begin{equation}   \label{eq:12}
f(\Delta X, \Delta P):=\Delta X \Delta P- \frac{1}{2}\left(1+\frac{\vert x\vert}{m_{pl}}(\Delta P)^2+\frac{\vert p\vert}{r_\Lambda}(\Delta X)^2\right)
\end{equation} 

then, we can find in analogy with \ref{eq:111} the minimal scales in position (UV cut-off) an momentum (IR cut-off) related to the concepts as they are described by an observer under the effects of Relative Locality and Co-locality. The minimal scale in position can be obtained with the conditions:

\begin{equation}   \label{eq:13}
\frac{\partial f}{\partial \Delta P}=0\;\;\;\;\;f(\Delta X, \Delta P)=0
\end{equation}

then:

\begin{equation}   \label{eq:14}
\Delta X_{min}=L\sqrt{\frac{q^2-1}{q^2}}\approx\sqrt{\frac{\vert x\vert}{m_{pl}}}
\end{equation} 

on the other hand, the minimal scale in momentum can be obtained from the general condition \ref{eq:113} applied to \ref{eq:12}. The result is:

\begin{equation}   \label{eq:15}
\Delta P_{min}=K\sqrt{\frac{q^2-1}{q^2}}\approx\sqrt{\frac{\vert p\vert}{r_\Lambda}}
\end{equation}

the results \ref{eq:14} and \ref{eq:15} are consistent with a q-Bargmann Fock algebra inside the $SU_q(n)$ in n dimensions, if we define the q-deformation parameter by:

\begin{equation}   \label{eq:16}
q\approx 1+\sqrt{\frac{\vert p\vert \vert x\vert}{r_\Lambda m_{pl}}}+...
\end{equation}

in agreement with the general constraint \ref{eq:28}. Eq. \ref{eq:11} is consistent with the original GUP formulation \ref{eq:10} and the constraint \ref{eq:28}. For deriving the result \ref{eq:16}, we have assumed that the condition:

\begin{equation}   \label{eq:17}
\vert x\vert \vert p\vert<<r_\Lambda m_{pl}
\end{equation}

is valid. Now we can derive the equivalent UV-IR mix scale when the effects of Relative locality and Relative co-locality are taken into account (together). The expression \ref{eq:124} is general. It represents the condition where the UV effects are equivalent to the IR ones. In the framework of Relative locality and Co-locality, the condition \ref{eq:124} is related to the scales at which the Relative locality effects cancel to the Relative Co-locality ones. We then need to obtain the new values for K and L already defined in \ref{eq:9} and \ref{eq:10}. From \ref{eq:16} and \ref{eq:17}, $q^2\approx 1+2\sqrt{\frac{\vert x\vert \vert p\vert}{r_\Lambda m_{pl}}}$; we can then replace this result in the general expression \ref{eq:10}:

\begin{equation}   \label{eq:18}
\Delta X\Delta P\geq \frac{1}{2}\left(1+\frac{1}{2}\sqrt{\frac{\vert p\vert \vert x\vert}{r_\Lambda m_{pl}}}\left(\frac{(\Delta X)^2}{L^2}+\frac{(\Delta P)^2}{K^2}\right)\right)
\end{equation}   

comparing this result with the expression \ref{eq:11}, we get:

\begin{equation}   \label{eq:19}
K=\left(\frac{1}{2}\right)^{1/2}\left(\frac{\vert p\vert}{\vert x\vert}\frac{m_{pl}}{r_\Lambda}\right)^{1/4}
\end{equation}

and:

\begin{equation}   \label{eq:20}
L=\left(\frac{1}{2}\right)^{1/2}\left(\frac{\vert x\vert}{\vert p\vert}\frac{r_\Lambda}{m_{pl}}\right)^{1/4}
\end{equation}

note that $KL\approx \frac{1}{2}$; consistent with \ref{eq:28}. If we replace these results in \ref{eq:124}, we get:

\begin{equation}   \label{eq:21}
\Delta X\approx \left(\frac{\vert x\vert}{\vert p\vert}\frac{r_\Lambda}{m_{pl}}\right)^{1/2}\Delta P
\end{equation}

this condition is obtained from the extremal condition \ref{eq:115} and \ref{eq:121} which are valid for any q-deformation parameter. Eq. \ref{eq:21} is the UV-IR mix condition under the Relative locality and co-locality effects. If we replace \ref{eq:21} in \ref{eq:11}, we then obtain the extended version of the UV-IR mix scale given by:

\begin{equation}   \label{eq:22}
\Delta P_{mix}\approx \left(\frac{1}{2}\right)^{1/2}\left(\frac{\vert p\vert m_{pl}}{\vert x\vert r_\Lambda}\right)^{1/4}
\end{equation} 

and:

\begin{equation}   \label{eq:23}
\Delta X_{mix}\approx \left(\frac{1}{2}\right)^{1/2}\left(\frac{\vert x\vert r_\Lambda}{\vert p\vert m_{pl} }\right)^{1/4}
\end{equation}

the results \ref{eq:22} and \ref{eq:23} are just extensions of the results \ref{eq:128} but including the effects of Relative locality and Relative co-locality. Note that if $\vert p\vert=\frac{1}{r_\Lambda}$ and $\vert x\vert=l_{pl}$, we recover the results \ref{eq:128} (with $\hbar=1$). Then UV and IR cut-offs will depend on the position and momentum which an observer would have with respect to the event.

\section{Isolated Relative Co-locality inside the $SU_q(n)$ deformed Heisenberg algebras}   \label{eq:New}

The previous section just showed the most general case where the observer is both, very far from the event and He/She is also moving with respect to it. Here I will introduce the $\Lambda$ scale as a fixed value in order to isolate the spatial curvature effects due to the non-commutative structure of momentum space. It is easy to demonstrate that we will recover the results of Section \ref{eq:dS BH} if we impose the condition $r_\Lambda\to\infty$. \\
If we want to isolate the spatial curvature effects, the observer's momentum relative to the source must reach its minimum value given by $p=\frac{1}{r_\Lambda}$; in such a case, the q-deformed parameter is just given by:

\begin{equation}   \label{eq:2333}
q\approx 1+\frac{1}{r_\Lambda}\sqrt{\frac{\vert x\vert}{m_{pl}}}
\end{equation}

note that if $\vert x\vert=l_{pl}$, then we recover the results of Section \ref{eq:8}. Here however, we are interested in the Relative locality regime. All the relevant results are just extensions of those obtained in the previous section. The minimum scale in position for example given by eq. \ref{eq:14}. However, the minimum scale in momentum is now given by the second equation of \ref{eq:30} (with $\hbar=1$). The values of the scale parameters K and L are just extensions of \ref{eq:19} and \ref{eq:20} and they are given by:

\begin{equation}   \label{eq:2334}
K\approx \left(\frac{1}{2}\right)^{1/2}\left(\frac{ m_{pl}}{\vert x\vert r_\Lambda^2}\right)^{1/4}
\end{equation} 

and:

\begin{equation}   \label{eq:2335}
L\approx \left(\frac{1}{2}\right)^{1/2}\left(\frac{\vert x\vert r_\Lambda^2}{m_{pl}}\right)^{1/4}
\end{equation}

additionally, the UV-IR mix scale is in this case, an extension of the results \ref{eq:22} and \ref{eq:23}. The new UV-IR mix scales are:

\begin{equation}   \label{eq:2336}
\Delta P_{mix}\approx \left(\frac{1}{2}\right)^{1/2}\left(\frac{m_{pl}}{\vert x\vert r_\Lambda^2}\right)^{1/4}
\end{equation}    

\begin{equation}   \label{eq:2337}
\Delta X_{mix}\approx \left(\frac{1}{2}\right)^{1/2}\left(\frac{\vert x\vert r_\Lambda^2}{m_{pl}}\right)^{1/4}
\end{equation}

The GUP expression in this case is:

\begin{equation}   \label{eq:2338}
\Delta X\Delta P\geq \frac{1}{2}\left(1+\frac{\vert x\vert}{m_{pl}}(\Delta P)^2+\frac{1}{r_\Lambda^2}(\Delta X)^2\right)
\end{equation}

which is consistent with the results of the previous sections.

\section{Conclusions}   \label{eq:C}

I have introduced the curvature effects inside the Relative Locality formalism as a non-commutative structure of momentum space in agreement with the an extended version of the Generalized Uncertainty Principle (GUP). The curvature effects manifest as a variation of the UV cut-off inside the q-Bargmann Fock scenario. As a consequence of this, there is a natural redshift effect for the observers located at a given distance relative to the event. This is a logical result in agreement with the standard notions of Gravitational Red-Shift generated by some spatial curvature.
If we try to localize a particle with a high energy photon for example at some point of the spacetime; for a distance observer this photon has a smaller frequency. For that observer the curvature effects inside the Relative locality formalism generate a natural red-shift effect consistent with an extended version of GUP. On the other hand, if the same observer is moving with some momentum relative to the event (photon), there is an additional blue-shift effect produced by the momentum curvature inside Relative Co-locality. The momentum curvature is introduced as a non-commutative effect of the spatial coordinates. We interpret the momentum curvature effects inside Relative Co-locality as a variation of the IR cut-off in a Generalized Uncertainty Principle. If the observer is at some distance from the event and he/she is also moving with respect to it, then the Relative locality and Co-locality effects compete each other. They become equally important at the extended UV-IR mix scales given by \ref{eq:22} and \ref{eq:23}. Where $\Delta X$ and $\Delta P$ are the scales of position and momentum of the event and p with x are the scales of position and momentum of the observer relative to the event.
It is possible to isolate the momentum or position curvature effects inside the q-Bargmann Fock formalism if we simply fix the position and/or momentum of the observer relative to the event. This automatically fixes the UV and/or IR cut-off scales in agreement with an extended version of GUP. \\     

{\bf Acknowledgement}

This work was inspired in the Topics of the Conference "Experimental Search for Quantum Gravity: The Hard Facts", organized at the {\bf Perimeter Institute for theoretical physics}.

\end{document}